\documentclass[3p,times,procedia]{elsarticle}
\usepackage{nupha_ecrc}
\usepackage{wrapfig}
\usepackage{placeins}
\usepackage{multicol}

\volume{00}

\firstpage{1}

\journalname{Nuclear Physics A}

\runauth{}

\jid{nupha}

\jnltitlelogo{Nuclear Physics A}

\usepackage{amssymb}

\usepackage[figuresright]{rotating}

\begin{document}

\begin{frontmatter}



\dochead{}

\title{High statistics study of in-medium S- and P-wave quarkonium states in lattice Non-relativistic QCD}


\author[a]{S.~Kim}
\author[b]{P.~Petreczky}
\author[c]{A.~Rothkopf\corref{cor1}$^{*,}$ }

\address[a]{Department of Physics, Sejong University, Seoul 143-747, Korea}
\address[b]{Physics Department, Brookhaven National Laboratory, Upton, NY 11973, USA}
\address[c]{Institute for Theoretical Physics,  Heidelberg University, Philosophenweg 16, 69120 Heidelberg, Germany}

\cortext[cor1]{Speaker}

\begin{abstract}
Many measurements of quarkonium suppression at the LHC, e.g.\ the nuclear modification factor $R_{AA}$ of $J/\Psi$, are well described by a multitude of different models. Thus pinpointing the underlying physics aspects is difficult and guidance based on first principles is needed.
Here we present the current status of our ongoing high precision study of in-medium spectral properties of both bottomonium and charmonium  based on NRQCD on the lattice. This effective field theory allows us to capture the physics of quarkonium without modeling assumptions in a thermal QCD medium. In our study a first principles and realistic description of the QCD medium is provided by state-of-the-art lattices of the HotQCD collaboration at almost physical pion mass. Our updated results corroborate a picture of sequential modification of states with respect to their vacuum binding energy. Using a novel low-gain variant of the Bayesian BR method for reconstructing spectral functions we find that remnant features of the Upsilon may survive up to $T\sim400$MeV, while the $\chi_b$ signal disappears around $T\sim270$MeV. The $c\bar{c}$ analysis hints at melting of $\chi_c$ below $T\sim190$MeV while some $J/\Psi$ remnant feature might survive up to $T\sim245$MeV. An improved understanding of the numerical artifacts in the Bayesian approach and the availability of increased statistics have made possible a first quantitative study of the in-medium ground state masses, which tend to lower values as $T$ increases, consistent with lattice potential based studies.
\end{abstract}


\end{frontmatter}

\vspace{0.5cm}

Heavy quarkonium, the bound states of a heavy quark-antiquark pair ($c\bar{c}$ charmonium, $b\bar{b}$ bottomonium) are versatile tools to investigate the different stages of a heavy-ion collisions \cite{Andronic:2015wma}. Dimuon spectra measured by CMS in both $p+p$ and $Pb+Pb$ collisions at LHC have revealed that $b\bar{b}$ suffers from a suppression of yields that is compatible with models of sequential melting \cite{Karsch:2005nk}. The underlying picture is that of a bound state produced in the early partonic stages, which traverses the collision center, hence samples the medium over an extended period of time and weakens in the process. On the other hand comparing charmonium yields at RHIC and LHC has shown a replenishment of $J/\Psi$ at high energies, which is well reproduced not only by transport models but also by fully thermal descriptions, such as the statistical model of hadronization \cite{BraunMunzinger:2000px}. This behavior interpreted in terms of regeneration hints at an at least partial kinetic equilibration of the charm quarks with their surroundings, an idea substantiated by the unambiguous signs of elliptic flow for the $J/\Psi$ particle measured by ALICE. In turn the observed charmonium may have lost substantial amounts of information on the intermediate evolution of the fireball and so promises to shed light predominantly on the latest stages of the quark-gluon plasma (QGP) just before hadronization. 

To provide first principles insight into this intricate phenomenology via the study of the spectral properties of heavy quarkonium is one of the central goals for theory \cite{Aarts:2016hap,Kim:2014iga, Aarts:2014cda}. One challenge lies in the fact that at the temperatures realized in the collision center, quarks and gluons are strongly correlated. This necessitates a fully non-perturbative approach. We will thus deploy lattice QCD simulations in imaginary time, which are able to capture the physics of the QCD medium from first principles even close to the crossover transition between QGP and hadronic phase. The description of the heavy quark d.o.f. on the other hand benefits from the separation of scales between the heavy quark mass ($m_c\simeq1.3$GeV, $m_b\simeq4.7$GeV) and the typical scales characterizing the QGP created in the collision $(T\sim\Lambda_{\rm QCD}\sim 200{\rm MeV})$. This forms the basis for an efficient theoretical description in terms of simplifying non-relativistic effective field theories. 

One such approach called pNRQCD consists of computing the static and complex valued in-medium potential between heavy quarks from first principles lattice QCD, to subsequently determine quarkonium spectral functions from solving an appropriate Schr\"odinger equation (see e.g.\ \cite{Burnier:2015tda}). A lot of intuition has been gained in this way. Melting of $Q\bar{Q}$ bound states is found to be a gradual process, i.e. spectral features do not disappear suddenly at a specific temperature. Therefore specifying a melting point is ambiguous. One possible definition is the temperature at which the in-medium binding energy equals the in-medium decay width of a state. Based on pNRQCD spectral functions, estimates of observables, such as the $\Psi'/J/\Psi$ ratio, have been obtained and are found to lie close to those from the statistical model. These computations, however, contain significant systematic uncertainties, since currently only the static contribution to the in-medium potential enters. Here we use a different EFT, which does not suffer from this source of uncertainty.

In lattice NRQCD the heavy quarks are described by non-relativistic Pauli spinor fields \cite{Lepage:1992tx}. These propagate in the background of the light thermal medium d.o.f., which themselves are provided by a conventional lattice simulation. This approach does not involve any modeling, as it is based on a systematic expansion of the QCD Lagrangian in terms of powers of the heavy quark velocity $v=\frac{\hat{p}_{\rm lat}}{m_Q a}$, where $a$ denotes the lattice spacing. The outcome of an NRQCD computation are quarkonium correlation functions at $T=0$ or $T>0$, projected to a specific physical quantum number channel (e.g. $b\bar{b}$: $^3S_1\equiv\Upsilon$,  $^3P_1\equiv \chi_{b}$) . 

The medium, in which the heavy quarks are immersed, is captured through state-of-the-art lattice simulations by the HotQCD collaboration \cite{Bazavov:2011nk}. These configurations incorporate the thermal physics of light $u,d$ and $s$ quarks with a pion mass of $m_\pi=161$MeV and a crossover temperature of $T_{\rm pc}=159$MeV almost at the physical point $T_{\rm pc}^{\rm cont}=154\pm9$MeV, and span a temperature range between $T\in[140-407]$MeV. The accuracy of NRQCD depends on the ratio $1/(m_Q n a)$ with $n$ the Lepage parameter, related to the temporal discretization of the theory (here $n_{b\bar{b}}=4$ and $n_{c\bar{c}}=8$). We consider the full $T$ range only for $b\bar{b}$ with $m_b a \in [2.759, 0.954]$, while for $c\bar{c}$ we restrict to the subset of $T\in[140-251]$MeV with $m_c a \in [0.757-0.42]$.

The determination of spectral functions from lattice NRQCD correlators represents an ill-posed inverse problem. The two quantities are related via an integral transform $D(\tau)=\int d\omega {\rm exp}[-\omega\tau]\rho(\omega)$. When discretized $D_i=\sum_l^{N_\omega} \Delta\omega_l {\rm exp}[-\omega_l\tau_i]\rho_l$,  with the number of datapoints $N_\tau$ much smaller than the number of  frequency bins $N_\tau<N_\omega$, it is clear that a $\chi^2$ fit of $\rho_l$ is underdetermined. Bayesian inference provides a mathematically solid path to give meaning to the inversion by incorporating additional, so called prior information. In essence the $\chi^2$ fitting functional is amended by a regulator, whose form is chosen to represent the prior information, e.g.\ the positive definiteness of the spectrum. Here we use one particular implementation of the Bayesian strategy, the BR method \cite{Burnier:2013nla}, which uses the regulator functional  $S_{\rm BR}=\int d\omega \big( 1-\rho/m+{\rm log}[\rho/m]\big)$. $m$ denotes the default model, which we set to a constant $m=1$, in order not to introduce artificial structure.

It has been found that in the presence of a small number of Euclidean datapoints $N_\tau\sim O(10)$ the BR method is susceptible to ringing \cite{Kim:2014iga}. We have reduced these artifacts here in two ways. On the one hand we base the reconstruction not only on the Euclidean correlator but also its Fourier transformed counterpart (denoted BRFT). The integral kernel connecting the Matsubara frequency data and the spectrum is significantly less damped and thus the data constraints the spectrum more strongly at high frequencies, where ringing artifacts predominantly appear. On the other hand we deploy a novel low-gain BR method, which contains an additional penalty on the arc length of the spectrum $S_{\rm BR}^{\rm smooth}=\int d\omega \big( (\partial_\omega \rho)^2+1-\rho/m+{\rm log}[\rho/m]\big)$. \setlength{\columnsep}{4pt}
\begin{wrapfigure}[36]{r}{0.4\textwidth}\vspace{-0.5cm}
  \begin{center}
     \includegraphics[scale=0.62]{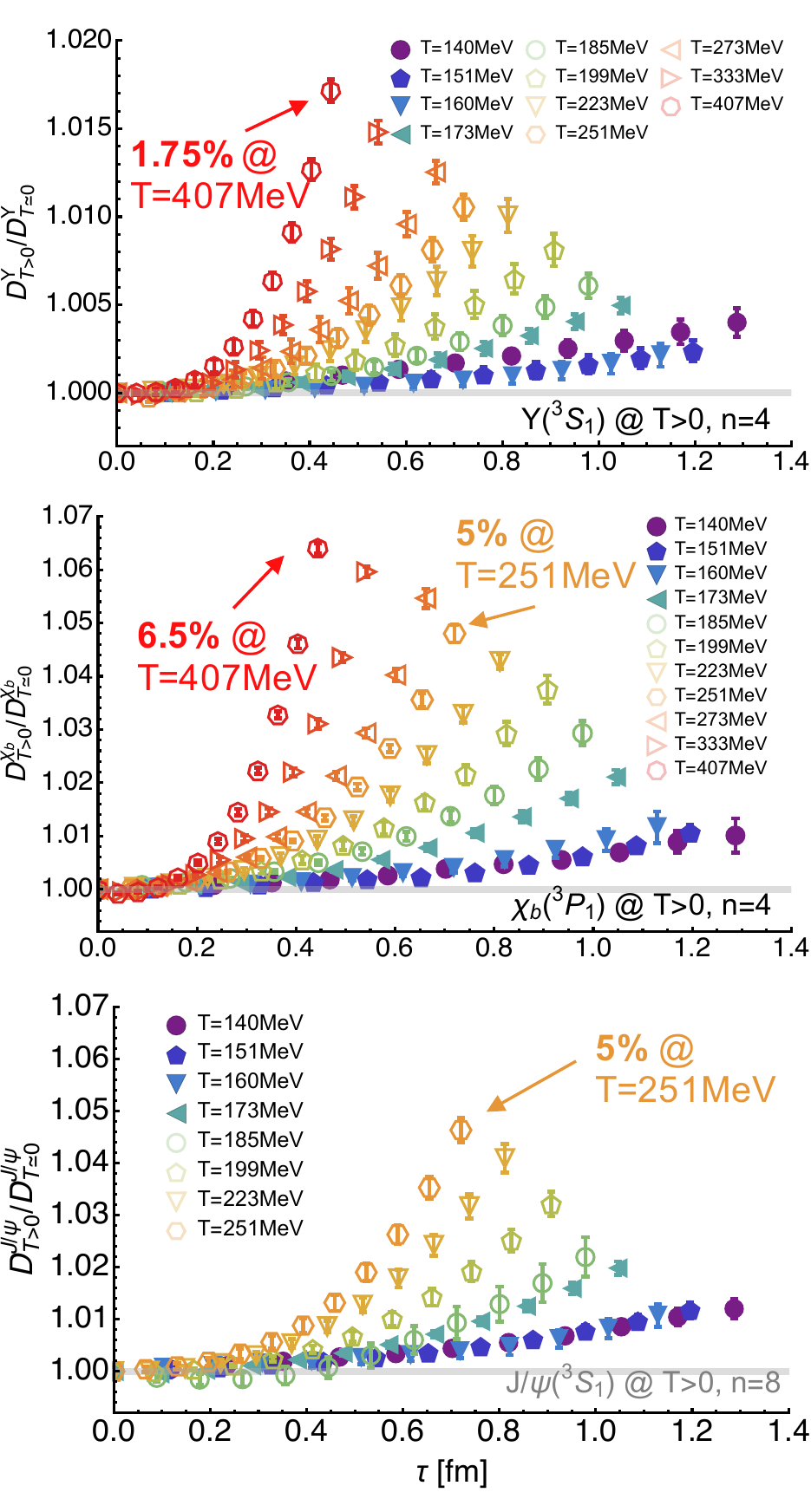}
  \end{center}\vspace{-0.7cm}
 \caption{Ratio of  the $Q\bar{Q}$ correlation functions at $T>0$ and $T\approx0$ in Euclidean time. In-medium modification emerges sequentially ordered according to the $E_{\rm bind}^{T=0}$. }\label{Fig:CompFiniteVac}
\end{wrapfigure}As we will show below it successfully suppresses artificial ringing e.g.\ in the reconstruction of non-interacting spectral functions and allows a more robust determination of the presence or absence of peaked spectral structures.

Before turning to in-medium spectra, we discuss first the ratio of the in-medium to the vacuum correlator shown in Fig.\ref{Fig:CompFiniteVac} for $\Upsilon$ (top), $\chi_b$ (center) and $J/\Psi$ (bottom). This quantity if different from unity, represent global in-medium modification. With significantly reduced statistical errors compared to \cite{Kim:2014iga} the previously indicated behavior is now corroborated. For $\Upsilon$ the correlator below $T_C$ starts off above unity, decreases slightly around $T_C$ and then goes over into a characteristic upward bend as temperature is further increased. Similar upward bending without the non-monotonous behavior around $T_C$ is observed for $\chi_b$ and $J/\Psi$ as well. Extending up to $T=407$ MeV we find that the absolute change in the correlators is still small, $1.75\%$ for $\Upsilon$ and $6.5\%$ for $\chi_b$. Consistent with expectations from the sequential suppression picture we confirm that states with similar binding energy, i.e. $\chi_b$ and $J/\Psi$ compared at the same temperature, here $T=251$MeV, show a very similar modification of $5\%$. 

One question of continued phenomenological interest are the melting temperatures of the individual quarkonium states. Since direct lattice studies are not yet able to resolve the in-medium excited states or the threshold, we instead focus on the disappearance of the bound state signal, which provides an upper limit to melting as defined above.  \setlength{\columnsep}{5pt}
\begin{wrapfigure}[28]{l}{0.35\textwidth}
  \begin{center}\vspace{-1.cm}
  \includegraphics[scale=0.56]{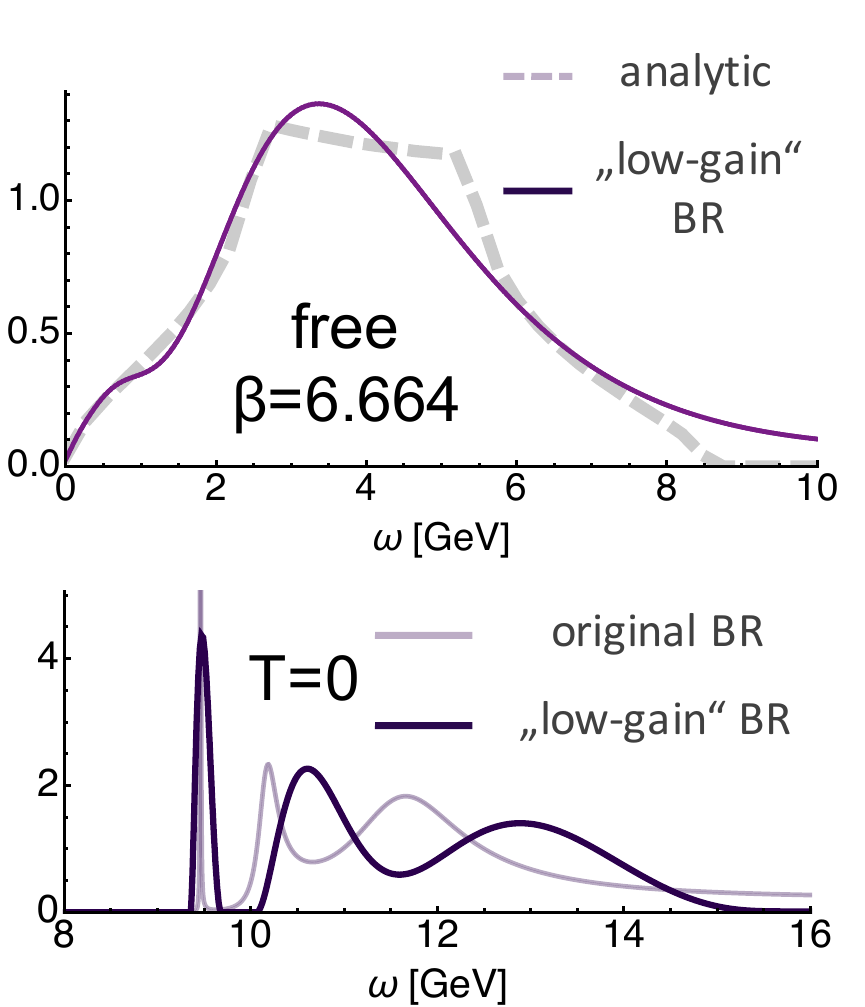}
  \end{center}\vspace{-0.6cm}
 \caption{(Top) Ringing free reconstruction of the $\beta=6.664$ free spectral function (gray dashed) with the low-gain BR method (dark solid). (Bottom) Comparison of the $T\approx0$ reconstruction at $\beta=6.664$ between the original (gray) and low-gain BR method (dark solid).}\label{Fig:CompTestLowGain}
 \end{wrapfigure}

Reconstructing spectra using the standard BR method may suffers from numerical ringing artifacts (c.f. Gibbs ringing) and thus wiggly features mimicking a bound state remnant can contaminate the spectrum and impede the determination of melting. The low-gain BR method improves this situation significantly. As seen in the top panel of Fig.\ref{Fig:CompTestLowGain}, if we deploy the method to a spectrum that is devoid of peaks, such as the noninteracting theory (gray dashed), the reconstruction (dark solid) also does not contain any artificial wiggly peaks. On the other hand if a genuine ground state peak is present, e.g.\ at $T\approx0$ the low-gain method is still able to pick up its position as seen in the lower panel.

Inspecting by eye the low-gain reconstructions at different temperatures we find that a remnant bound state signal for $\Upsilon$ is present up to the highest temperature $T=407$MeV. The $\chi_b$ state with its much smaller vacuum binding energy on the other hand disappears already from the spectrum below $T=273$MeV. For $J/\Psi$ where we only explore up to $T=251$MeV we find a signal to be present, whereas $\chi_c$ is gone at $T=185$MeV. These values are consistent with our previous findings, which were based on the less reliable comparison between the reconstructed free and interacting spectral functions. In case that we are have ascertained that a bound state peak persists in the spectrum, we then use the original BR method to extract its position. 

In order to obtain more quantitative insight, we need to thoroughly understand the sources of systematic error in the Bayesian reconstruction.  \setlength{\columnsep}{4.5pt}
\begin{wrapfigure}[40]{r}{0.4\textwidth}
  \begin{center}\vspace{-0.4cm}
\includegraphics[scale=0.53]{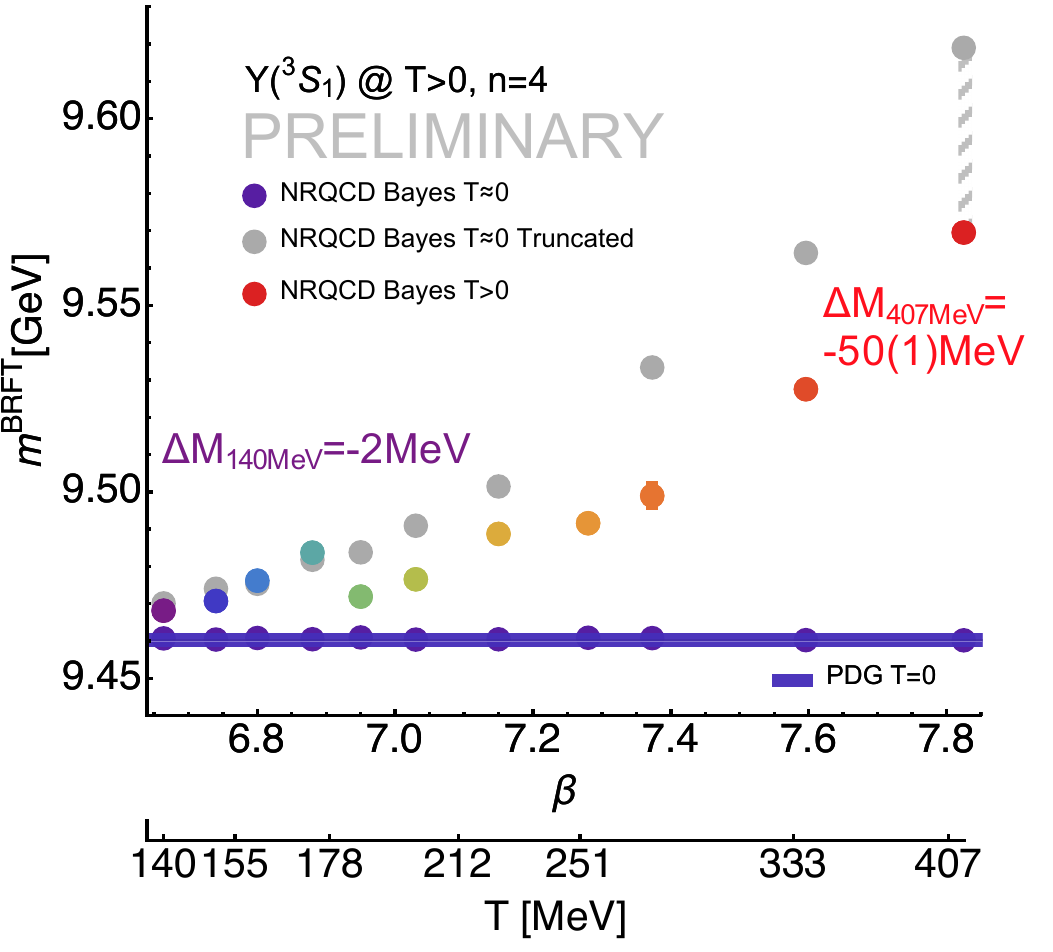}
\includegraphics[scale=0.53]{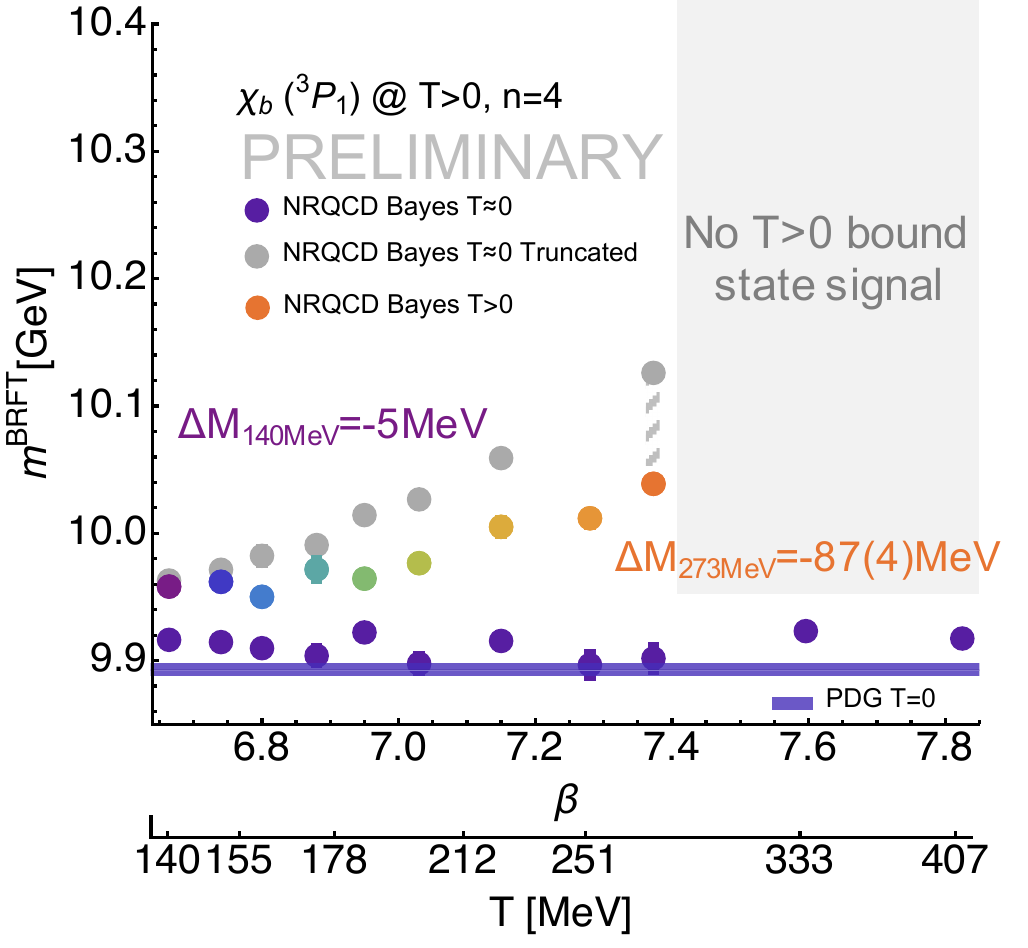}
\includegraphics[scale=0.53]{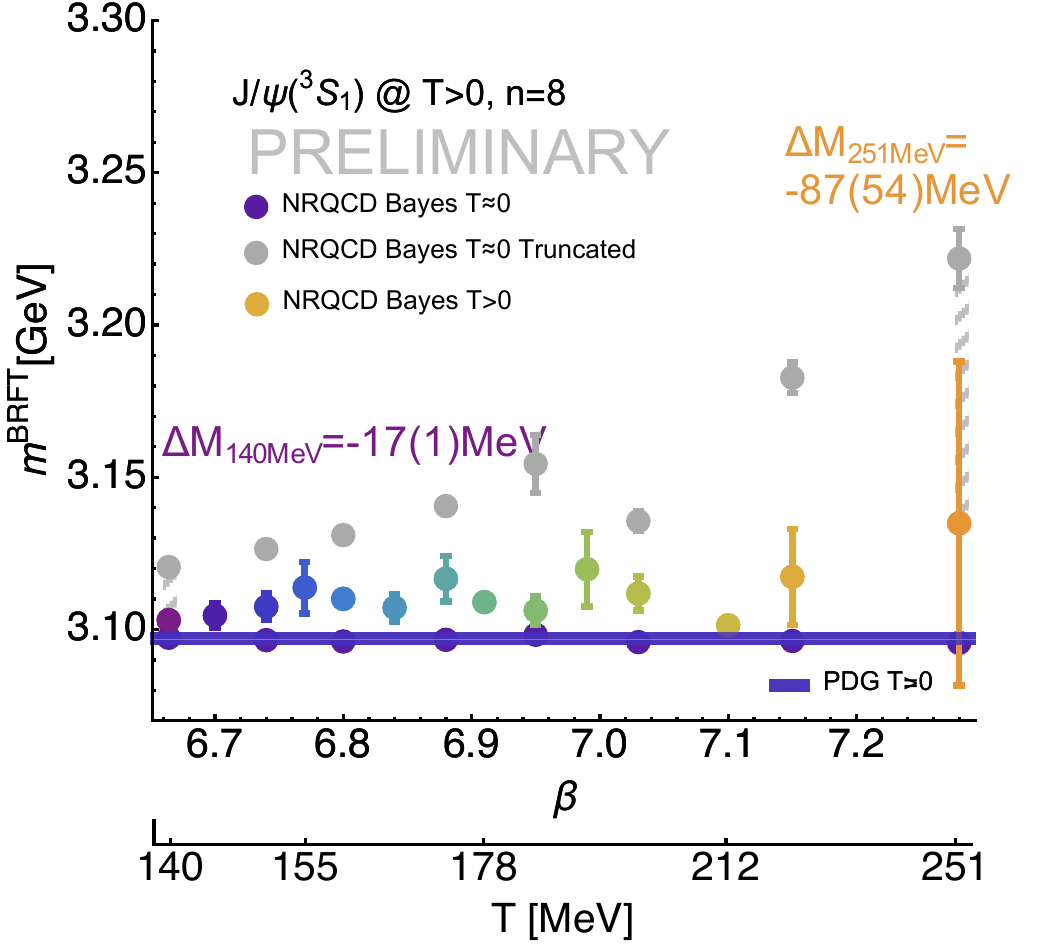}
  \end{center}\vspace{-0.6cm}
 \caption{Preliminary in-medium ground state masses (colored points) for $\Upsilon$ (top), $\chi_b$ (center) and $J/\Psi$ (bottom). The best estimates at $T\approx0$ are given as dark blue points, while the light gray points denote those from the truncated $T\approx0$ data. In-medium effects are manifest in the difference between colored and gray points. (Note the different temperature scale for $J/\Psi$)}\label{Fig:QuantPeakPos}
\end{wrapfigure}As discussed in \cite{Kim:2014iga} the finite number of points and the limited physical extent in Euclidean time are the main factors affecting reconstruction quality at different $T$. We thus carry out the following crosscheck: the correlator at $T\approx0$ ($\tau_{\rm max}=32-64$) is truncated to twelve \newpage \noindent points, the same number available at $T>0$. Feeding this reduced dataset to the BR methods yields a reconstruction of the ground state with broadened features, systematically shifted to higher frequencies. In Fig.\ref{Fig:QuantPeakPos} this effect on the mass for $\Upsilon$, $J/\Psi$ and $\chi_b$ is manifest in the difference between the best possible reconstruction (dark blue points) and the truncated one (light gray points). The latter then represent the actual $T\approx0$ baseline to which the in-medium masses have to be compared to.

Now if we carry out the reconstruction for the actual $T>0$ correlator sets, we obtain the ground state masses shown in Fig.\ref{Fig:QuantPeakPos} as colored points. Note that a naive comparison to the $T\approx0$ mass would have resulted in the conclusion that masses increase with temperature. However according to the appropriate $T\approx0$ baseline, we observe that for $\Upsilon$ there is no significant change in the mass up to $T=173$MeV, above which it shows a clear offset towards lower values, though the effect is small. Even at $T=407$MeV the difference is only $\Delta M=-50(1)$MeV, small compared to the vacuum binding energy of $1.1$GeV. For $\chi_b$ with a smaller vacuum binding energy $0.6$GeV, in-medium effects appear to set in at lower $T$ and are also stronger e.g.\ $\Delta M=-87(4)$MeV at $T=273$MeV. Remember that in contrast to $\Upsilon$ we do not find a clear bound state signal remnant for $\chi_b$ above $T=273$MeV. 

In accordance with the correlator ratios for charmonium we expect similar behavior as for $\chi_b$. Since the amount of statistics for $c\bar{c}$ is lower its results are however less precise. The $J/\Psi$ in-medium mass already changes by $\Delta M=-17(1)$ at $T=141$MeV, which grows to $\Delta M=-87(54)$MeV at $T=251$MeV. 

Both our updated correlators, as well as the spectral reconstructions based on the BR method corroborate the picture of sequential in-medium modification of quarkonium w.r.t. its vacuum binding energy, indicated in \cite{Kim:2014iga}. Increasing the statistics for $c\bar{c}$ in order to reach the same quantitative level of precision for $\Delta M$ is work in progress.

SK is supported by Korean NRF grant No.\  2015R1A2\\ A2A01005916,  PP by the U.S. DOE under contract DE-SC0012704. This work is part of and supported by the DFG Collaborative Research Centre "SFB 1225 (ISOQUANT)". We thank HotQCD for providing the gauge configurations for this study.

\end{document}